\documentstyle[12pt]{article}
\newcommand{\be}{\begin{equation}}
\newcommand{\ee}{\end{equation}}
\newcommand{\ba}{\begin{eqnarray}}
\newcommand{\ea}{\end{eqnarray}}

\begin{document}
\begin{center}
 {\bf\Large{
   Metafluid dynamics and Hamilton-Jacobi formalism}}
\end{center}
\begin{center}
{\bf Dumitru Baleanu}\footnote[1]{ E-mails:
dumitru@cankaya.edu.tr, baleanu@venus.nipne.ro}
\end{center}
\begin{center}
Department of Mathematics and Computer Science, Faculty of Arts
and Sciences, Cankaya University-06530, Ankara, Turkey
\end{center}

\begin{center}
and
\end{center}

\begin{center}
Institute of Space Sciences, P.O BOX, MG-23, R 76900
Magurele-Bucharest, Romania
\end{center}

\begin{abstract}

Metafluid dynamics was investigated within Hamilton-Jacobi
formalism and the existence of the hidden gauge symmetry was
analyzed. The obtained results are in agreement with those of
Faddeev-Jackiw approach.

\end{abstract}



\section{Introduction}

    The hydrodynamic turbulence represents an ancient important issue from theoretical
and experimental point of view. Recently, this problem was
subjected to an intense debate because there are many phenomena
which are turbulent, for example in astrophysics, cosmology,
biomechanics, meteorology [1, 2, 3]. Recently, an alternative
description for the Clebsch decomposition of currents in fluid
mechanics was proposed [4] and its non- abelian extensions were
obtained [5].The hidden gauge invariance was studied in Dirac and
extended frame formalism [6].
 A new approach for investigation of the fluid
turbulence was proposed recently [7].The method named the
metafluid dynamics, is based on the use of the analogy between
Maxwell electromagnetism and turbulent hydrodynamics and it
describes the dynamical behavior of average flow quantities in
incompressible fluids flows having high Reynolds numbers in a
similar way as it was done to get the macroscopic electromagnetic
fields [8]. The average procedure was obtained using the spatial
filtering method indicated in [9]. A Lagrangian for the metafluid
dynamics was proposed recently in [10].The theory was analyzed for
the first time as a constrained system from the symplectic point
of view and a hidden gauge symmetry was reported [10]. Therefore,
a new procedure describing the gauge symmetries of the constrained
systems should be applied to the same Lagrangian. Hamilton-Jacobi
(HJ) formalism, based on $Carath\acute{e}odory's$ equivalent
Lagrangians method [11] is an alternative method for quantization
of constrained systems [12]. The action provided by (HJ) is useful
for the path integral quantization method of the constrained
systems. It was proved that the integrability conditions of (HJ)
formalism and Dirac's consistency conditions [13] are equivalent
[14] and the equivalence of the chain method [15] and (HJ)
formalism was performed [16].

The main aim of this paper is to investigate the metafluid
dynamics within (HJ) formalism and to give the corresponding
justification for the existence of its hidden symmetry.

The plan of the paper is as follows:

 In Sec. 2 the (HJ) formalism
is briefly introduced. In Sec. 3 the metafluid dynamics is
presented and its treatment within symplectic formalism is briefly
reviewed. In Sec. 4 the (HJ) formulation of metafluid dynamics was
investigated. Sec.5 is dedicated to our conclusions.

\section{Hamilton-Jacobi formalism}

The basic idea of this new approach [12] is to consider the
constraints as "Hamiltonians" and to involve all of them in the
process of finding the action. Let us assume that a given
degenerate Lagrangian L admits the following primary
"Hamiltonians"

\be H_{\alpha}^{'}=p_{\alpha}+H_{\alpha}(t_\beta,q_a,p_a). \ee

We define the canonical Hamiltonian $H_0$ as

\be H_0=-L(t,q_i,\dot{q_\nu},\dot{q_a}=w_a)+p_aw_a +{\dot
q_\mu}p_\mu;  \nu=0,n-r+1,\cdots,n.
 \ee

 The equations of motion are obtained as total differential
 equations in many variables as given below:

 \begin{equation}
 dq_a=\frac{\partial H_{\alpha}^{'}}{\partial p_a}dt_{\alpha},
 dp_a=-\frac{\partial H_{\alpha}^{'}}{\partial q_a}dt_\alpha,
 dp_\mu=-\frac{\partial H_{\alpha}^{'}}{\partial t_\mu}dt_\alpha,
 \mu=1,\cdots,r.
 \end{equation}

 and the (HJ) function is given by

 \be
 dz=(-H_\alpha +p_\alpha\frac{\partial H_{\alpha}^{'}}{\partial
 p_a})dt_\alpha
 \ee

The set of equations (3) is integrable if and only if [12]

\be [H_\alpha^{'},H_\beta^{'}]=0,\forall\alpha,\beta. \ee

The method is straightforward for constrained systems [17] having
finite degree of freedom [18] but it becomes, in some cases, quite
difficult to be used for field theories. The main difficulty comes
from the fact that some surface terms may play an important role
in closing the algebra of "Hamiltonians" but some of them have no
physical meaning from the (HJ) point of view. Another problem is
the treatment of the second-class constraints systems within (HJ)
formalism . In this particular case the "Hamiltonians" are not in
involution and it is not a unique way to solve this problem [19,
20, 21].

\section{Metafluid dynamics}

Based on the analogy between Maxwell electromagnetism and
turbulent hydrodynamics Marmanis (see for more details Ref. [7]
and the references therein) proposed an approximative theory such
that the equations describing the dynamic variables are linear but
the nonlinearities emerge as sources of turbulent motion. Marmanis
constructed a system of equations containing the vorticity
$\overrightarrow{\varpi}=\nabla\times\overrightarrow{u}$ and the
Lamb vector
$\overrightarrow{l}=\overrightarrow{w}\times\overrightarrow{u}$ as
follows

\be \nabla.\overrightarrow{w}=0,\frac{\partial
\overrightarrow{w}}{\partial
t}=-\nabla\times\overrightarrow{l}+\upsilon\nabla^2\overrightarrow{w},\nabla.\overrightarrow{l}=n(\overrightarrow{x},t)
\ee

\be \frac{\partial \overrightarrow{l}}{\partial t}=u^2
\nabla\times\overrightarrow{w}-\overrightarrow{j}(\overrightarrow{x},t)+\upsilon{\nabla
n(\overrightarrow{x},t)}-\upsilon\nabla^2\overrightarrow{l}, \ee
where the turbulent current is given as

\be \overrightarrow{j}(\overrightarrow{x},t)=\overrightarrow{u}n
+\nabla\times(\overrightarrow{u}.\overrightarrow{w})\overrightarrow{u}+\overrightarrow{w}\times\bigtriangledown(\Phi+\overrightarrow{u}^2)+2(\overrightarrow{l}.\nabla)\overrightarrow{u},
\ee and the turbulent charge $n(\overrightarrow{x},t)$ has the
following form \be
n(\overrightarrow{x},t)=-\nabla^2\Phi(\overrightarrow{x},t). \ee
In (9) the Bernoulli energy function $\Phi(\overrightarrow{x},t)$
has the expression

\be \Phi(\overrightarrow{x},t)=\frac{p}{\rho}
+\frac{\overrightarrow{u}^2}{2}, \ee

where $p(\overrightarrow{x},t)$ is the pressure , $\rho$ is the
density, $\overrightarrow{u}(\overrightarrow{x},t)$ represents the
velocity field and v is the kinematic viscosity.

Taking the averaging process of (6) and (7) we obtain (see for
more details [7]) \be
\nabla.\overrightarrow{\varpi}=0,\frac{\partial\overrightarrow{\varpi}}{\partial
t}=-\nabla\times\overrightarrow{l}+\upsilon\nabla^2\overrightarrow{\varpi},\nabla.\overrightarrow{l}=n(\overrightarrow{x},t),
\ee

\be \frac{\overrightarrow{\partial l}}{\partial t}=c^2
\nabla\times\overrightarrow{\varpi}-\overrightarrow{J}(\overrightarrow{x},t)+\upsilon\nabla
n(\overrightarrow{x},t)-\upsilon\nabla^2 \overrightarrow{l},\ee

where

\be
\overrightarrow{\varpi}=<\overrightarrow{\varpi}>,\overrightarrow{l}=<\overrightarrow{l}>,\overrightarrow{J}=<\overrightarrow{j}>,
c^2=<u^2>,\overrightarrow{u}=<\overrightarrow{u}>, \ee and

\be \phi=<\Phi>,n(\overrightarrow{x},t)=<n(\overrightarrow{x},t)>.
\ee

Since the Lagrangian density of the classical electromagnetism [8]
is given by the very well known expression

\be L=\frac{1}{2}(\overrightarrow{E^2}-\overrightarrow{B^2}),
 \ee

 the analogy between electromagnetism and turbulence allow us to
 write the Lagrangian density turbulence [10] as

 \be
 L=\frac{1}{2}(\overrightarrow{l}^2-c^2\overrightarrow{w}^2).
 \ee
Inserting the expressions of $\overrightarrow{l}$ and
$\overrightarrow{w}$ into (16) the form of the Lagrangian density
becomes \be L=\frac{1}{2}(\nabla\phi-\frac{\partial
\overrightarrow{u}}{\partial t}+ v\nabla^2\overrightarrow{u})^2
-\frac{1}{2}c^2(\nabla\times\overrightarrow{u})^2. \ee

In [10] the authors considered the case when sources are not
zero.The interaction Lagrangian density

\be
L_{int}=\overrightarrow{J}.\overrightarrow{u}-n\phi-v\overrightarrow{u}.\nabla
n. \ee

was added to (17) and the total Lagrangian density corresponding
to the metafluid dynamics can be written as

\be L= \frac{1}{2}(\nabla\phi-\frac{\partial
\overrightarrow{u}}{\partial t}+v\nabla^2\overrightarrow{u})^2-
\frac{1}{2}c^2(\nabla\times\overrightarrow{u})^2+\overrightarrow{J}.\overrightarrow{u}-n\phi-v\overrightarrow{u}.\nabla
n. \ee

Using the Faddeev-Jackiw analysis [22] the set of constraints
corresponding to (19) was obtained [10] as

\be \pi^0=0,\nabla.\overrightarrow{\pi}+n=0. \ee

Imposing $\phi$ being a constant and using the condition of
incompressible of fluid $\nabla.\overrightarrow{u}=0$, the Dirac's
brackets among the space space fields were calculated as

\be \{u_i(\overrightarrow{x}),\pi_i(\overrightarrow{y})\}_{DB}=
(\delta_{ij}-\frac{\partial_i^x\partial_j^y}{\nabla^2})\delta(\overrightarrow{x}-\overrightarrow{y}).
 \ee

 Finally, it was proved that the Lagrangian (19) admits the
 following gauge symmetry [10]

 \be
 \delta u_i=\partial_i\epsilon,\delta\pi_i=0,\delta\phi=-{\dot\epsilon}.
 \ee

 \section{Hamilton-Jacobi analysis}
 The starting point in (HJ) is the degenerate Lagrangian density given by (19).
The canonical momentum conjugate to $\overrightarrow{u}$ has the
form \be
\overrightarrow{\pi}(\overrightarrow{x},t)=-\overrightarrow{l}(\overrightarrow{x},t),
 \ee
 but
 \be
 \pi^0=0.
 \ee

Here $\pi^0$ is the canonical momenta associated to $\phi$.Making
use of (19), (23) and (24) the canonical Hamiltonian density is
given by

\be
H_c=\frac{1}{2}\overrightarrow{\pi}^2-\overrightarrow{\pi}.\nabla\phi
+\frac{1}{2}c^2(\nabla\times \overrightarrow{u})^2
+\upsilon\overrightarrow{\pi}.\nabla^2\overrightarrow{u}-\overrightarrow{u}.\overrightarrow{J}+\phi
n +\upsilon\overrightarrow{u}.\nabla n \ee

The Hamiltonians densities to start with are

\be H_0^{'}=p_0+H_c,H_1^{'}=\pi^0. \ee

The equations of motion corresponding to (26) have the following
expressions

\be
d\overrightarrow{u}=(\overrightarrow{\pi}+\upsilon\nabla^2\overrightarrow{u}-\nabla\phi)d\tau,
 \ee

 \be
d\overrightarrow{\pi}=(\overrightarrow{J}-\upsilon\nabla n
-c^2\nabla\times(\nabla\times\overrightarrow{u})-\upsilon\nabla^2\overrightarrow{\pi})d\tau
 \ee
The next step is to impose the integrability conditions and to
make the above system integrable. Imposing $dH_1^{'}=0$ we obtain
another "Hamiltonian" as

\be H_2=\nabla.\overrightarrow{\pi} +n \ee

Using the fact that $n=-\nabla^2\phi$ and making zero the
variation of (29) obtain \be H_3=(\nabla.\overrightarrow{J}+
2\upsilon\nabla^2n +\dot{n})d\tau.
 \ee

Analyzing the form of (30) we observed that the process of finding
new "Hamiltonians" finished. The gauge variable Á is determined
from $H_3 = 0$ but the problem of imposing the gauge fixing
$\nabla.\overrightarrow{u}=0$ inside of (HJ) formalism remains not
yet justified. At the first sight, the set of "Hamiltonian"
densities \be H_0^{''}=p_0+\frac{1}{2}\overrightarrow{\pi}^2
+\frac{1}{2}c^2(\nabla\times\overrightarrow{u})^2
+\upsilon\overrightarrow{\pi}.\nabla^2\overrightarrow{u}-\overrightarrow{u}.\overrightarrow{J}
+\upsilon\overrightarrow{u}.\nabla n,
H_1^{'}=\pi^0,H_2=\nabla.\overrightarrow{\pi} +n,
 \ee

 are in involution but $H_2$ is not in the form required in
(1). We have two options at this stage: to use an extended
phase-space or to make a canonical transformation such that $H_2$
becomes a momentum. Knowing that all Poisson brackets are
canonical invariants [23] we conclude that it is possible to
perform a canonical
 transformation for metafluid theory. In this way the existence of the hidden gauge
symmetry is justified inside of (HJ) formalism.

In addition, we observed that in the inertial range $\eta\ll
InertialRange \ll L$ (see Ref. [2] for notations and more details)
the constraints obtained in [10] by using the symplectic analysis
are the same as those produced by (HJ) formalism. By inspection,
we can easily check the "Hamiltonians" $H_0^{"},H_1^{'}$ and $H_2$
produce the same Dirac's brackets as obtained in the symplectic
formalism [10] but in (HJ) approach we are working on the original
phase space extended with the pair $(p_0,\tau)$.

  \section{Conclusion}

 In (HJ) one of the main problem is to put all "Hamiltonians" in the form given by
(1). The analysis of the metafluid dynamics is one example of
theory possessing a gauge symmetry but one "Hamiltonian" is not in
the requested form, so it creates problems in managing the hidden
gauge symmetry. We observed that it is possible to perform a
canonical transformation such that $H_2$ becomes a momentum and
the "Hamiltonians" remain in involution. In this manner we argue
the existence of the gauge hidden symmetry of metafluid dynamics
within (HJ) formalism. In [18] and very recently in [24] the
important role of the canonical transformations for (HJ) analysis
of second-class constrained systems was discussed. In this paper
we claimed that the canonical transformations are needed and for
some of the first class constrained systems. In addition, we found
that the surface terms played an important role in finding the
total differential equations (27) and (28). It was reported that
(HJ) and Faddeev-Jackiw approaches gave the same set of
constraints although the formalisms have complectly different
structures and different mechanisms of identified the gauge
variables.

 \section {Acknowledgments}
This work is partially supported by the Scientific and Technical
Research Council of Turkey.


\begin{thebibliography}{99}

\bibitem{1}
L.D. Landau and E. M. Lifshits: {\it Fluid Mechanics} (Pergamon
Press, Oxford, 1980).

\bibitem{2}
U. Frisch: Turbulence: {\it The Legacy of A. N. Kolmogorov},
Cambridge University Press, Cambridge (1995).

\bibitem{3}
S. Weinberg: {\it Gravitation and Cosmology} (J. Wiley, 1972).

\bibitem{4}
T.S. Nyawelo, J.W. van Holten, and S. G. Nibbelink: {\it Phys.
Rev.} {\bf D68} (2003) 125006,\\

\bibitem{5}
B. Bistrovic, R. Jackiw, H. Li, V.P. Nair and S.-Y. Pi: {\it
Phys.Rev.}{\bf D67} (2003) 025013.

\bibitem{6}
S. Ghosh,arXiv:hep-th/0107190,\\
 S.Ghosh, {\it J.Phys.} {\bf A35} (2002) 10747.

\bibitem{7}
H. Marmanis: {\it Phys. Fluids} {\bf 10} (1998) 1428; PhD
thesis:Analogy between the Electromagnetic and Hydrodynamic
Equations: Application to Turbulence,
2000,http://www.cfm.brown.edu/people/marmanis.

\bibitem{8}
J. D. Jackson: {\it Classical Electrodynamics} (J. Willey, New
York, 1983).

\bibitem{9}
 G. Russakov: {\it Am. J. Phys.} {\bf 38} (1970) 1188.

\bibitem{10}
A. C. R. Mendes, C. Neves, W. Oliveira, and F. I. Takakura: {\it
Braz. Journ. Phys.} {\bf vol. 33 no. 2} (2003) 346.

\bibitem{11}
C. $Carath\acute{e}odory$: {\it Calculus of Variations and Partial
Differential Equations of the First Order} Part II, (Holden-Day ,
1967).

\bibitem{12}
Y. $G\ddot{u}ler$: {\it Nuovo Cimento} {\bf B107} (1992) 1143.

\bibitem{13}
P.A. M. Dirac: Lectures on Quantum Mechanics (Yeshiva Univ. Press,
New York, 1967).

\bibitem{14}
B.M. Pimentel, R.G. Teixteira and J.L. Tomazelli: {\it Ann.Phys.}
{\bf 267}(1998) 75.

\bibitem{15}
P. Mitra and R. Rajaraman: {\it Ann. Phys.} {\bf 203} (1990) 137;
{\it Ann. Phys.} {\bf 203} (1990) 157.

\bibitem{16}
D. Baleanu and Y. $G\ddot{u}ler$: {\it Nuovo Cimento} {\bf B118
no.6} (2003) 615.

\bibitem{17}
M. Henneaux and C. Teitelboim: Quantization of Gauge Systems
(Princeton Univ. Press, Princeton, N. J. 1992).

\bibitem{18}
D. Baleanu and Y. $G\ddot{u}ler$: {\it Nuovo Cimento} {\bf  B115
no.3} (2000) 319.

\bibitem{19}
D. Baleanu and Y. $G\ddot{u}ler$ : {\it J. Phys. A: Math. Gen.}
{\bf  34} (2001) 73.

\bibitem{20}
D. Baleanu and Y. $G\ddot{u}ler$: {\it Mod. Phys. Lett. B} {\bf
vol 16 no.13} (2001) 873.

\bibitem{21}
D. Baleanu and Y. $G\ddot{u}ler$: {\it Int. Journ. Mod. Phys. A}
{\bf vol 16 no. 13} (2001) 2391.

\bibitem{22}
L. Faddeev and R. Jackiw: {\it Phys. Rev. Lett.} {\bf 60} (1988)
1692.

\bibitem{23}
H. Goldstein, C. Poole and J. Safko: (Classical Mechanics, Third
Edition, Addison Wesley, 2002).
\bibitem{24}
K. D. Rothe and F.G. Scholtz, {\it Ann. Phys.}, {\bf 308},(2003)
639.
\end{thebibliography}
\end{document}